\documentclass[11pt]{article}
\usepackage{times}
\usepackage{sidecap}
\usepackage{geometry}
\usepackage{floatrow}
\usepackage{float}
\usepackage{hyperref}
\usepackage{graphics,graphicx}
\usepackage{subcaption}
\usepackage{epsf,wrapfig}

\usepackage{natbib,bm}

\usepackage{paralist}
\let\olditem\item
\renewenvironment{thebibliography}[1]{%
  \section*{\refname}
  \let\par\relax
  \renewcommand{\item}[1][]{\olditem[\textbullet]}%
  \inparaenum}{\endinparaenum}

\usepackage{enumitem}

\newcounter{aff}

\DeclareGraphicsExtensions{.jpg,.pdf,.png,.eps}

\usepackage{caption}
\usepackage{xcolor}

\usepackage{tcolorbox}

\bibpunct{(}{)}{;}{a}{}{,}

\usepackage{soul}

\definecolor{darkcyan}{rgb}{0.0, 0.0, 0.7}

\definecolor{boxcolor}{RGB}{193, 229, 255}
\definecolor{figboxcolor}{RGB}{193, 229, 255}
\definecolor{boxcolorgray}{RGB}{227, 227, 227}
\definecolor{darkcyan}{rgb}{0.0, 0.0, 0.7}

\sethlcolor{boxcolorgray}

\def\gtrsim{\lower 2pt \hbox{$\, \buildrel {\scriptstyle >}\over
{\scriptstyle \sim}\,$}}
\def\lesssim{\lower 2pt \hbox{$\, \buildrel {\scriptstyle <}\over
{\scriptstyle \sim}\,$}}

\def\approxlt{\lower.2em\hbox{$\buildrel < \over \sim$}}
\def\approxgt{\lower.2em\hbox{$\buildrel > \over \sim$}}

\newcommand{\kms}{km s$^{-1}$}
\newcommand{\myr}{M$_{\odot}$ yr$^{-1}$}
\newcommand{\msun}{M$_{\odot}$}

\newcommand{\ci}{[C\,{\sc i}]}
\newcommand{\oi}{[O\,{\sc i}]}
\newcommand{\cii}{[C\,{\sc ii}]}
\newcommand{\hi}{H\,{\sc i}}

\def\be{\begin{equation}}
\def\ee{\end{equation}}
\def\bea{\begin{eqnarray*}}
\def\eea{\end{eqnarray*}}

\def\jref@jnl#1{{\rm#1}}
\def\aj{\jref@jnl{AJ}}			
\def\araa{\jref@jnl{ARA\&A}}		
\def\apj{\jref@jnl{ApJ}}			
\def\apjl{\jref@jnl{ApJ}}		
\def\apjs{\jref@jnl{ApJS}}		
\def\ao{\jref@jnl{Appl.~Opt.}}		
\def\apss{\jref@jnl{Ap\&SS}}		
\def\aap{\jref@jnl{A\&A}}		
\def\aapr{\jref@jnl{A\&A~Rev.}}		
\def\aaps{\jref@jnl{A\&AS}}		
\def\azh{\jref@jnl{AZh}}			
\def\baas{\jref@jnl{BAAS}}		
\def\jrasc{\jref@jnl{JRASC}}		
\def\memras{\jref@jnl{MmRAS}}		
\def\mnras{\jref@jnl{MNRAS}}		
\def\pra{\jref@jnl{Phys.~Rev.~A}}	
\def\prb{\jref@jnl{Phys.~Rev.~B}}	
\def\prc{\jref@jnl{Phys.~Rev.~C}}	
\def\prd{\jref@jnl{Phys.~Rev.~D}}	
\def\pre{\jref@jnl{Phys.~Rev.~E}}	
\def\prl{\jref@jnl{Phys.~Rev.~Lett.}}	
\def\pasp{\jref@jnl{PASP}}		
\def\pasj{\jref@jnl{PASJ}}		
\def\qjras{\jref@jnl{QJRAS}}		
\def\skytel{\jref@jnl{S\&T}}		
\def\solphys{\jref@jnl{Sol.~Phys.}}	
\def\sovast{\jref@jnl{Soviet~Ast.}}	
\def\ssr{\jref@jnl{Space~Sci.~Rev.}}	
\def\zap{\jref@jnl{ZAp}}			
\def\nat{\jref@jnl{Nature}}		
\def\iaucirc{\jref@jnl{IAU~Circ.}}
\def\aplett{\jref@jnl{Astrophys.~Lett.}}
\def\apspr{\jref@jnl{Astrophys.~Space~Phys.~Res.}}
\def\bain{\jref@jnl{Bull.~Astron.~Inst.~Netherlands}}
\def\fcp{\jref@jnl{Fund.~Cosmic~Phys.}}
\def\gca{\jref@jnl{Geochim.~Cosmochim.~Acta}}
\def\grl{\jref@jnl{Geophys.~Res.~Lett.}}
\def\jcp{\jref@jnl{J.~Chem.~Phys.}}	
\def\jgr{\jref@jnl{J.~Geophys.~Res.}}	
\def\jqsrt{\jref@jnl{J.~Quant.~Spec.~Radiat.~Transf.}}
\def\memsai{\jref@jnl{Mem.~Soc.~Astron.~Italiana}}
\def\nphysa{\jref@jnl{Nucl.~Phys.~A}}
\def\physrep{\jref@jnl{Phys.~Rep.}}
\def\physscr{\jref@jnl{Phys.~Scr}}
\def\planss{\jref@jnl{Planet.~Space~Sci.}}
\def\procspie{\jref@jnl{Proc.~SPIE}}

\def\ion#1#2{#1$\;${\small\rm\@Roman{#2}}\relax}

\newcount\lecurrentfam
\def\LaTeX{\lecurrentfam=\the\fam \leavevmode L\raise.42ex
\hbox{$\fam\lecurrentfam\scriptstyle\kern-.3em A$}\kern-.15em\TeX}
\def\sizrpt{
(\fontname\the\font): em=\the\fontdimen6\font, ex=\the\fontdimen5\font
\typeout{
(\fontname\the\font): em=\the\fontdimen6\font, ex=\the\fontdimen5\font
}}
\definecolor{dark}{rgb}{ 0.34, 0.09, 0.49}

\usepackage{fancyhdr,url}
\usepackage{transparent}
\usepackage{tablefootnote}
\usepackage{mathpazo}

\begin{document}

\noindent
\begin{raggedright}
{\Huge\bfseries Resolving the Assembly of Disk Galaxies from the Epoch of Reionisation to Cosmic Noon \\[0.5em] 
}\end{raggedright}

\noindent

\noindent Melanie Kaasinen$^{1*}$, Mahsa Kohandel$^{2,3*}$, Francesca Rizzo$^{4*}$, Cecilia Bacchini$^{5*}$, Anita Zanella$^2$, Lucie Rowland$^6$, Miguel Querejeta$^7$, Hannah R. Stacey$^8$, Andrea Lapi$^{9}$, Umberto Maio$^{10}$, Kirsty M. Butler$^{11}$, Gareth Jones$^{12}$, Emily Wisnioski$^1$,  Justin Spilker$^{13}$, Livia Vallini$^{2}$, Andrea Pallottini$^{14}$, Bel\'{e}n Alcade Pampliega$^{15}$, Juan Molina$^{16,17}$, Pavel Mancera Pi\~{n}a$^{7}$, 
Corentin Cadiou$^{18}$, Giulia Despali$^{2}$, Francesco Salvestrini$^{10}$, Federico Lelli$^{19}$, %
Daizhong Liu$^{20,21}$, Maru\u{s}a Brada\u{c}$^{22}$, Karina Caputi$^{4}$, Michele Ginolfi$^{23}$, Matteo Messa$^{2}$, Isabella Lamperti$^{23}$, Marcel Neeleman$^{24}$, Alba Vidal-Garc\'{i}a$^{25}$, Jeremy Fensch$^{26}$, Seiji Fujimoto$^{27}$, Jacqueline A. Hodge$^{6}$,  %
Gerg\"{o} Popping$^{8}$,  Sabine Thater$^{28}$, Elisabete da Cunha$^{29}$, Leindert A. Boogaard$^{6}$, Manuel Aravena$^{30,17}$, Takafumi Tsukui$^{31}$, Deanne Fisher$^{32}$, %
James McGregor$^{1}$,Pierre Cox$^{33}$, Jonathan Freundlich$^{34}$,  Mamta Pandey-Pommier$^{26}$,  Miguel Pereira-Santaella$^{35}$, Aristeidis Amvrosiadis$^{36}$,  Vanshika Kansal$^{32}$     

\vspace{1em}
{
\footnotesize{
\noindent %
$^1$ Research School of Astronomy and Astrophysics, Australian National University, Canberra, ACT 2611, Australia, \\ 
$^2$ Istituto Nazionale di Astrofisica (INAF), Via Gobetti 93/3, 40129 Bologna, Italy \\ 
$^3$ Scuola Normale Superiore, Piazza dei Cavalieri 7, I-56126 Pisa, Italy, \\ 
$^4$ Kapteyn Astronomical Institute, University of Groningen, Landleven 12, 9747 AD, Groningen, The Netherlands, \\ 
$^5$ DARK, Niels Bohr Institute, University of Copenhagen, Jagtvej 155, 2200 Copenhagen, Denmark \\ 
$^*$ \emph{The first four authors contributed equally to this white paper} \\
All other affiliations can be found at the end of the paper.
}
}
\begin{tcolorbox}[boxsep=3pt,left=0pt,right=0pt,top=1pt,bottom=1pt,colframe=boxcolor,colback=boxcolor]
\justify
\textbf{Abstract:} 
        Most of the mass and angular momentum of today's disk galaxies assembled at $z>1$. Yet, it remains unclear how these galaxies acquired their gas, how their baryons and dark matter coevolved, and how feedback and radial gas transport shaped their disks. Addressing these gaps requires resolved observations of cold gas, which carries most of the mass and angular momentum involved in galaxy assembly. By resolving CO and \cii\ at kpc scales, ALMA revealed a surprising population of rotation-dominated disks, even at $z>2$, exhibiting kinematic signatures consistent with spiral arms, bars, and tidal tails. Yet, these studies are limited to massive ($M_\star>10^{10.5}$\,\msun) and highly star-forming (SFRs$\gtrsim100$\,\myr) galaxies, the progenitors of today's most massive elliptical and disk galaxies. To understand how most present-day spirals (like the Milky Way) assembled, it is crucial to resolve their low-mass progenitors, which dominate the star-forming galaxy population by number at $z>1$. Achieving this goal demands a tenfold increase in line sensitivity and hence a next-generation sub-/millimetre facility: ALMA2040. This leap in sensitivity will enable routine resolved studies of hundreds of $M_\star>10^9$\,\msun\ galaxies, including those on and below the main sequence of star-forming galaxies, providing representative, statistical galaxy samples with sub-kpc constraints on the cold gas kinematics and morphology. By uniquely providing resolved CO, \ci, and \cii\ observations across $z=1-6$, ALMA2040 will trace a much larger fraction of the cold-gas reservoir than is accessible today, extending resolved studies from rare massive and/or lensed systems to representative galaxy populations. In this way, ALMA2040 will transform our understanding of how galaxy disks formed and evolved. 
\end{tcolorbox}

\newpage{}

\section{Introduction}

   Galaxy assembly is governed by the flow of baryons through dark matter (DM) halos. Gas accreted from the cosmic web acquires angular momentum, settles into rotationally supported disks, and 
   promotes star formation. Throughout their assembly, DM halos shape the gravitational potential in which galaxies evolve, whereas baryonic processes such as accretion, feedback, and 
   radial gas transport alter the distribution of mass and angular momentum 
   \citep{SnD_2015,NnO_2017}. Together, these processes regulate key galaxy properties, including size, stellar mass, and star formation rate (SFR). Yet understanding how these processes interact and collectively govern galaxy evolution 
   remains a major challenge in modern astrophysics. 

  The epoch at $z>1$ spans two key phases of galaxy evolution: the initial assembly of galaxies at $z>3$ and their subsequent growth and dynamical transformation at Cosmic Noon ($z\sim1-3$). Yet despite major observational progress at $z>1$, several fundamental questions remain unanswered.
    \begin{itemize}[leftmargin=*, nosep]
        \item How do galaxies acquire gas and angular momentum?
        \item When do dynamically cold disks become a common and stable outcome of galaxy assembly? 
        \item How do baryons and DM halos coevolve, establishing the dynamical scaling relations, dark matter fractions, and DM halo structures observed in galaxies today? 
        \item Once disks are in place, how do gravitational instabilities, feedback, and non-axisymmetric structures 
        redistribute mass and angular momentum within galaxies? 
    \end{itemize}
    Answering these questions requires resolved (sub-kpc scale) observations of cold (molecular and atomic) gas, which carries most of the mass and angular momentum involved in disk assembly. Crucially, such observations must extend to the low-mass galaxies that dominate the population by number and represent the progenitors of today's $L_\star$ spiral galaxies, including the Milky Way. 

    Sub-/millimetre observations of cold-gas tracers such as CO, \ci, \cii, and \oi\ provide a powerful means of probing gas accretion, disk dynamics, and angular-momentum transport at $z>1$. These cold gas tracers directly probe the turbulence and radial flows that regulate star formation in disks, complementary to ionised-gas tracers, which more sensitively probe energetic processes like outflows \citep{ejdetjarn_2022,kohandel_2024}. For dusty, gas-rich systems, cold-gas tracers provide a unique window into heavily obscured regions where ionised-gas tracers like H$\alpha$ are strongly attenuated, enabling kinematic studies across the full extent of the gas disk. Resolved observations of the main atomic gas tracer, \hi, remain inaccessible beyond the local Universe; even with the increased sensitivity of the Square Kilometre Array (SKA), such studies will be limited to massive galaxies at $z\lesssim1$ \citep{lelli_2026}. Thus, sub-/millimetre observations provide a critical window on the gas reservoirs and kinematics that govern disk assembly at $z>1$.

    \begin{figure*}[t!]
      \includegraphics[width=\textwidth,trim={0cm 0.3cm 0cm 0.1cm},clip]{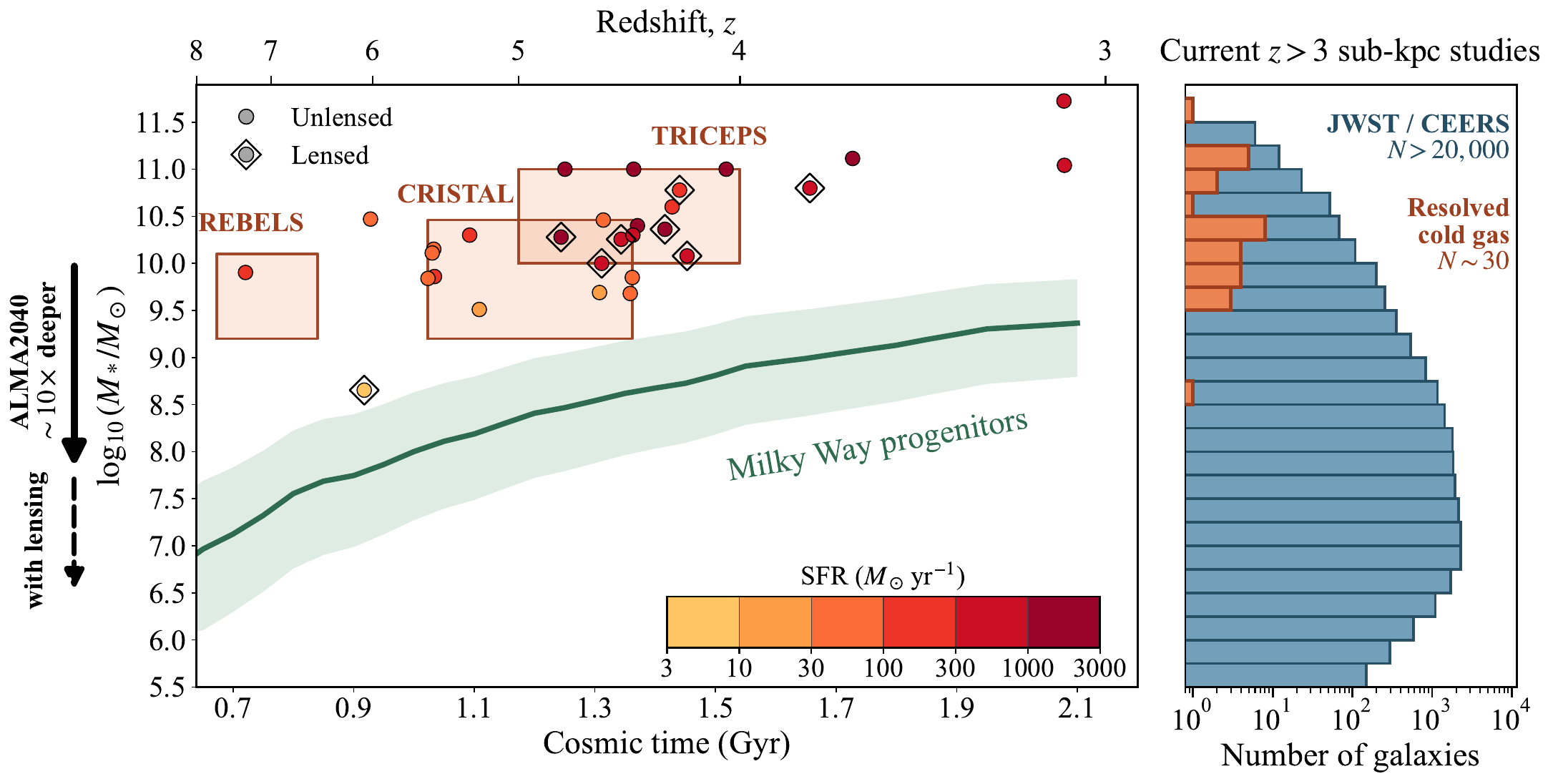}
       \caption{Current sub-/millimetre facilities have resolved the most massive and highly star-forming galaxies, shown here for $z>3$, leading to significantly smaller samples with resolved cold gas observations than resolved stellar studies (right panel). By resolving progenitors of local spiral galaxies like the Milky Way \citep[green line:][]{Cadiou2026} in $\lesssim10$ hours each (vs the $1000$\, hours required with ALMA), ALMA2040 will enable statistical resolved studies that are representative of the star-forming galaxy population, matching the population already resolved by JWST.}
       \label{fig:parameter_space}
   \end{figure*}
    
    Over the last 15 years, ALMA has revolutionised studies of disk assembly at $z>1$. By resolving cold gas at kpc scales, it has revealed a population of massive, highly star-forming galaxies that are more rotation-dominated than expected from ionised gas observations or simulations \citep{neeleman_2020,lelli_2021,roman-oliveira_2023,lelli_2023,rizzo_2024}. Several of these systems also exhibit evidence of spiral arms, bars, and tidal interactions \citep{hodge_2019,tsukui_2021,amvrosiadis_2025,kaasinen_2026}. Extending these studies beyond rare starbursts to representative samples of massive main-sequence galaxies has required several large program-scale investments. For example, the large program ``CONDOR'' has dedicated 230\,hours of ALMA time to resolve the gas kinematics and surface-density distributions of just $\sim10$ massive (M$_\star>10^{10.5}$\,\msun) main-sequence galaxies at $z=1-3$. Together with the high-mass parameter space covered at $z>3$ (Fig.~\ref{fig:parameter_space}), this highlights that ALMA lacks the sensitivity required to resolve the dominant population of lower-mass galaxies. 

\begin{figure}[t!]
        \centering
        \includegraphics[width=0.9\textwidth,trim={0cm 0.2cm 0cm 0cm},clip]{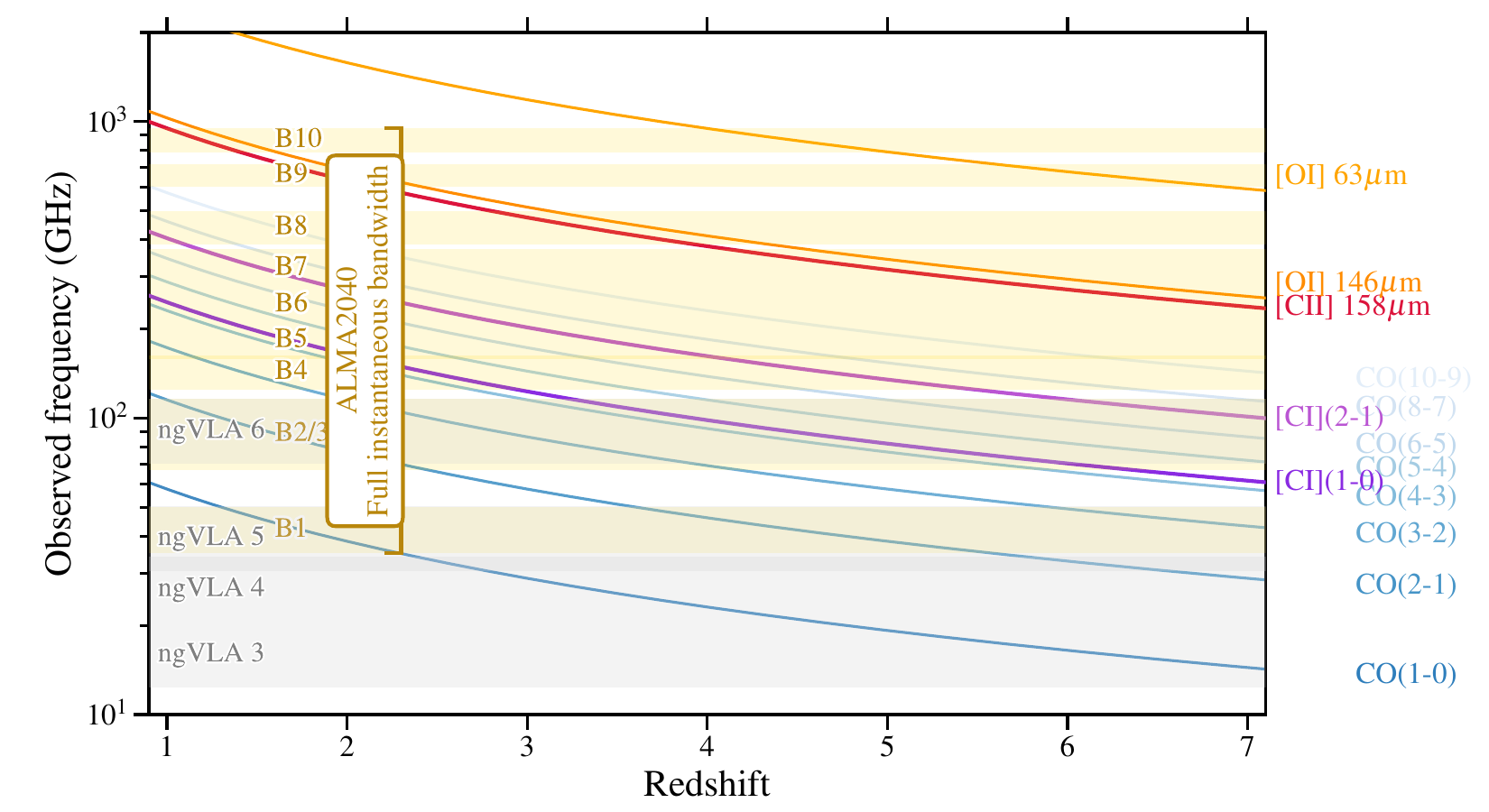}
            \caption{ALMA2040's \emph{instantaneous} spectral coverage. Multiple cold-gas transitions will be observed simultaneously, thereby providing an unprecedented view of the dynamics of distant galaxies. At $z\gtrsim1$, ALMA2040 will uniquely provide access to \ci\ and \cii, which are critical to resolving low-mass galaxies not accessible through CO observations with the ngVLA.}
            \label{fig:alma2040_v_ngvla}
    \end{figure}

    A complete view of disk assembly requires CO, \ci, and \cii\ to be mapped jointly, as their relative distributions and kinematics reveal how mass and angular momentum are partitioned between different phases of the cold interstellar medium (ISM) (Sec.~\ref{sec:tech_reqs}). Mapping \ci\ and \cii\ becomes increasingly critical at low masses, as they are expected to remain detectable over a larger fraction of the disk than CO. However, with its current sensitivity, ALMA takes several tens of hours to resolve \ci\ or \cii\ at $z\lesssim3$, even for the most massive galaxies, limiting such studies to a handful of extreme and/or lensed galaxies \citep{lamarche_2018,rooney_2025}. Although ALMA's unique site and frequency coverage make it the only facility capable of covering CO, \ci, and \cii\ across $z=1-6$, jointly observing all three still requires separate time-intensive observing programs. Hence, most multi-line studies are limited to fortuitous line combinations covered by the same receiver \citep{tadaki_2019,gomez_2025,kaasinen_2026}. A next-generation sub-/millimetre interferometer with $10\times$ the line sensitivity of ALMA and instantaneous coverage of all receiver bands (Fig.~\ref{fig:alma2040_v_ngvla}) is therefore essential to resolve the key processes that fuel and shape the dominant population of galaxies, from $z\sim6\to1$. \emph{This facility is ALMA2040.}
    
    ALMA2040 would play a groundbreaking role in the next-generation facility landscape (Sec.~\ref{sec:synergies}). At $z=1-3$, ALMA2040 will provide a crucial complement to the matched, high-resolution views of stars and ionised gas in typical galaxies that will be provided by the Extremely Large Telescope (ELT) and Giant Magellan Telescope (GMT). Beyond $z\sim3$, it will provide the primary means of resolving the dynamics and angular momentum of typical low-mass galaxies. ALMA2040 will also extend the ngVLA's view of diffuse molecular gas, uniquely resolving the high-frequency tracers that ALMA has shown to be critical for probing the kinematics of galaxy disks to large radii. By uniquely combining the sensitivity, angular resolution, and high-frequency spectral coverage needed to resolve the cold-gas morphology and kinematics of the low-mass galaxy population at $z>1$, ALMA2040 will provide the missing cold-gas perspective needed to connect gas accretion, galaxy dynamics, and disk growth from the Epoch of Reionisation to Cosmic Noon.

\section{Scientific context and open questions}
    \label{sec:motivation}

    \subsection*{Case 1: The feeding and settling of disks}
        \label{sub:feeding}

        Simulations and analytic models predict that typical $z\gtrsim2$ galaxies are fueled mainly by cold-mode accretion via filaments, with hot-mode accretion from galactic coronae becoming increasingly important at higher halo masses and later times \citep{keres_2005,dekel_birnboim_2006,dekel_2009_coldstreams,vandeVoort_2011}. Coherent, high-angular-momentum accretion is expected to build extended, rotationally supported disks, whereas discrete, misaligned, and/or low-angular-momentum accretion builds more compact and dynamically hotter galaxies \citep{dekel_burkert_2014,zolotov_2015}. Mergers and interactions also increase turbulence, and/or drive gas flows, although gas-rich systems may rapidly reform disks post-encounter. Some simulations further indicate that dynamically cold disks are transient structures that exist only while gas accretion is coherent and coplanar \citep{kretschmer_2022}. The relative impact of these processes is expected to depend on halo mass, with more massive halos predicted to host more stable, rotationally supported disks \citep{el_badry_2018,sultan_2026}. Thus, current models  make distinct predictions for the sizes, rotational support, angular momentum, and lifetimes of disk structures in early galaxies.  Testing these predictions requires resolved measurements of the cold-gas kinematics and surface density across large galaxy samples spanning a wide range of masses, environments, and redshifts. A handful of ALMA studies have tested these predictions, finding high rotational support, specific angular momentum, and baryon-to-DM mass ratios that challenge standard models of cold-gas accretion \citep{rizzo_2026,kaasinen_2026}. Yet, these studies are limited to the most massive starburst galaxies at $z>2$: for which simulations predict the earliest emergence of stable disks. 

\begin{figure}[t!]
           \centering    
           \vspace{-0.2cm}
           \includegraphics[width=\textwidth]{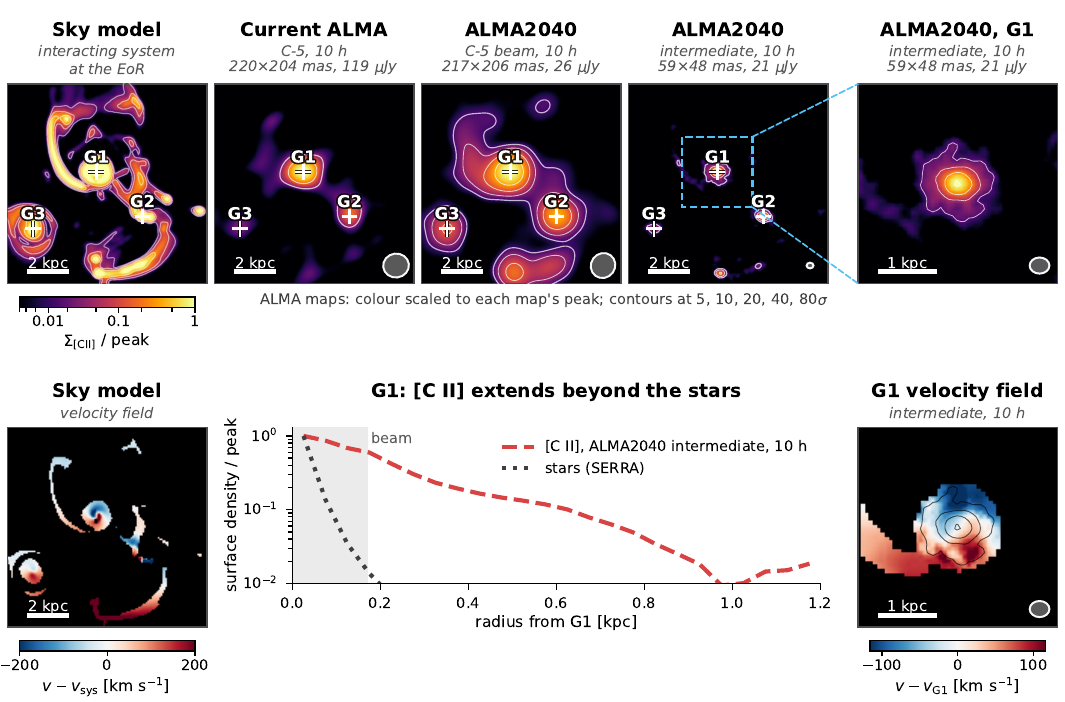}
            \caption{\textbf{ALMA2040 resolves interacting galaxies, tidal bridges, and extended velocity fields at
            $z=6.3$.} \textit{Top row: } Moment-0 maps of the \cii\ emission from the SERRA model galaxy (left); and different mock observations (labelled at top). \textit{Bottom, left and right:} Moment-1 map of input model and extended-configuration observation of G1. \textit{Right middle:} The recovered \cii\ surface density profile of G1 (red-dashed line) extends well beyond the stellar component (black dotted line).} 
        \label{fig:merger-vs-disk}
    \end{figure}

        Interactions are also pivotal in shaping early galaxies. ALMA and JWST are increasingly revealing tidal bridges, extended gas structures, and disturbed kinematics around $z=4-6$ galaxies, indicating ongoing exchanges of mass and angular momentum between neighbouring systems \citep{ginolfi_2020b,di_cesare_2024,solimano_2024,posses_2025}. Rather than representing a simple dichotomy between mergers and disks, these observations point toward a complex picture in which interactions may perturb, reshape, or even help build disks. However, most ALMA observations lack the angular resolution and surface-brightness sensitivity needed to distinguish individual galaxy components within interacting systems or trace the faint tidal structures through which mass and angular momentum are exchanged (e.g. Fig.~\ref{fig:merger-vs-disk}). As a result, mergers, clumpy disks, and interacting systems are difficult to disentangle, particularly at $z\gtrsim5$. This begs the question: how efficiently do interactions redistribute gas and angular momentum, and how important are these processes for the evolution of typical galaxies at $z\gtrsim2$? By resolving both the internal kinematics of galaxy disks and the faint tidal bridges and gas flows connecting interacting systems, ALMA2040 will place these processes in a statistical context and establish how galaxy interactions contribute to the growth and settling of disks across cosmic time.

    \subsection*{Case 2: The coevolution of galaxies and their dark matter halos}
        \label{sub:dm_baryon_coevolution}
    
            The coupled evolution of baryons and DM is a central problem in galaxy formation. At $z\sim0$, disk galaxies obey tight scaling relations linking the mass, rotation, and specific angular momentum of their baryons to the depth of the gravitational potential and spin of their host DM halos, 
            indicating a remarkably tight baryon-DM coupling \citep{mcgaugh_2000,lelli2016,mancera_pina_2021,mancera_pina_2025,geesink_2025}. Yet, it remains unclear how and when this coupling emerged, how it evolved in the early stages of galaxy assembly, and whether it depends on environment. At the same time, the fundamental nature of DM remains unresolved. By the 2040s, any viable DM model matching the wealth of constraints from facilities such as Euclid, Rubin, and LISA in the nearby Universe, must also explain the evolution of DM halo structure across cosmic time. The $z\sim6$ Universe is a 
            powerful testing ground: galaxies are observed much closer to the epoch at which their halos formed, where different DM models are predicted to diverge most \citep{correa_2025, mezghanni_2026}. In this regime, the evolution of dynamical scaling relations and the internal structure of DM halos become direct tests of CDM against alternative DM models. Recovering this information requires spatially resolved measurements of the circular velocity, obtained from the combination of rotation velocity ($V_\mathrm{rot}$) and pressure support traced by the gas velocity dispersion ($\sigma$), together with independent constraints on the stellar and gas mass distributions.

            Current and upcoming facilities leave a critical gap in the measurements needed to connect baryons and DM across cosmic time -- the ability to trace cold gas to large radii. Wide-area surveys from Euclid and Rubin will characterise the large-scale environments in which galaxies evolve, while JWST, the ELT, and GMT will deliver spatially resolved stellar mass distributions for representative samples at $z>1$. Among currently accessible gas tracers, \cii\ offers the most promising route to measuring outer rotation curves at $z>1$ (Sec.~\ref{sec:tech_reqs}). Yet even the highest-quality ALMA observations cannot detect \cii\ emission to sufficiently large radii to constrain the DM distribution directly \citep{lelli_2021,roman_oliveira_2024,rizzo_2026}. ALMA2040 will extend \cii\ measurements to the radii required to isolate DM contributions, while simultaneous observations of CO and \ci\ at $z<4$ (\ci\ and \oi\ at $z>4$) will constrain the gas mass distribution. Together, these capabilities will deliver the first statistical measurements of the evolution of dynamical scaling relations and DM density profiles at $z>1$, revealing how baryon-DM coupling depends on mass, environment, and cosmic time, and providing a decisive test of CDM against alternatives models (e.g., SIDM and WDM).

    \subsection*{Case 3: How feedback and gas flows build disk structure}
        \label{sub:feedback_flows}
    
        Star formation and disk evolution are governed by the interplay between gravity and stellar feedback. Gravitational instabilities can drive radial gas flows and angular momentum transport, while stellar feedback injects energy and momentum into the ISM, maintaining turbulent support against collapse and thereby regulating star formation. However, the relative importance of these processes for galaxy evolution remains unknown. Nearby galaxies are broadly consistent with models in which stellar feedback is the dominant source of turbulence \citep{ostriker_kim_2022,bacchini_2020}. At Cosmic Noon, however, the high velocity dispersions inferred from ionised gas \citep{wisnioski_2015} and large UV/H$\alpha$ clumps initially motivated a picture in which violent disk instabilities (VDIs) drive substantial radial transport and contribute significantly to the turbulent energy budget \citep{krumholz_2018}. In this VDI scenario, gas-rich disks become marginally unstable, fragmenting into massive star-forming clumps that migrate through the disk while redistributing mass and angular momentum. More recent ALMA observations of cold gas have instead revealed lower velocity dispersions and more stable disks than inferred from ionised-gas tracers, raising the possibility that stellar feedback is also the dominant driver of turbulence at $z=1-3$ \citep{bacchini_2024,rizzo_2024,roman_oliveira_2024}. But these measurements remain limited to a few tens of massive ($>10^{10.5}$\,\msun) galaxies, leaving the origin of turbulence and the prevalence of gravitational instabilities in the broader galaxy population poorly constrained. 

        Beyond their global dynamical state, galaxies develop substructures that reshape their evolution by redistributing mass and angular momentum. Bars, spiral arms, tidal tails, and clumps generate non-axisymmetric gravitational torques that drive radial gas flows and regulate where gas accumulates, influencing the growth of central structures and stellar disks \citep{dekel_2022,meidt_2024}. JWST has revealed substantial populations of barred and spiral galaxies out to $z\gtrsim3$, demonstrating that these substructures emerged earlier than previously thought \citep{guo_2025,es_2025}. However, there are few measurements of the dynamical impact of these structures. Recent large-program observations of massive main-sequence galaxies \citep{pastras_2025,jolly_2026} and targeted observations of extreme starbursts \citep{amvrosiadis_2025,kaasinen_2026} at $z\sim1-3$ have begun to resolve the kinematic signatures of such features. However, for most studies, non-circular motions remain unresolved and are absorbed into measured line widths, inflating the inferred velocity dispersions. Routinely resolving gas streaming across bars, spiral arms, and tidally perturbed disks will enable robust measurements of the true turbulence across galaxy disks, while separating the rates of radial gas transport and angular momentum redistribution. Such measurements require the tenfold leap in line sensitivity provided by ALMA2040, thereby enabling statistical tests of how stellar feedback and gravitational instabilities regulate the turbulent energy budget of high-redshift galaxies.

\section{ALMA2040: Technical Requirements}
    \label{sec:tech_reqs}

    The three science cases in Sec.~\ref{sec:motivation} place the following requirements on the sensitivity, angular and spectral resolution, and bandwidth. 
    \begin{description}
        \item[Case 1] requires sub-kpc measurements of the cold-gas kinematics ($V_\mathrm{rot}(R)$, $\sigma(R)$, and inflows) and mass distribution ($\Sigma_\mathrm{gas}(R)$).
        Achieving this demands sufficient sensitivity to detect cold gas, 
        sufficient angular resolution to sample the radial kinematic profiles, 
        and sufficient spectral resolution to recover the intrinsic gas turbulence. It also requires sensitivity to low-surface-brightness gas associated with tidal bridges, tails, and ongoing mass exchange. 
        \item[Case 2] relies on the accurate recovery of both the rotation curve and gas mass distribution. 
        This requires sufficiently high resolution to measure the inner mass distribution and sufficiently high surface-brightness sensitivity to detect faint \cii\ emission at large radii. \item [Case 3] builds on the kinematic requirements of case 1, additionally requiring the spatial and spectral resolution to identify non-axisymmetric structures (such as spiral arms and bars) and measure the gas flows through which they redistribute mass and angular momentum. This demands sufficient angular resolution to resolve the characteristic scales of these structures (a few 100 pc) and sufficient sensitivity to robustly discern non-circular motions.
    \end{description}    

    All three science cases require observations of multiple cold-gas tracers. \cii\ provides the most promising route to recovering kinematics at large radii. Unlike rest-optical tracers (which primarily trace the star-forming disk), C$^+$ 
    remains abundant in diffuse atomic gas and the CO-dark molecular disk. Hence, it is critical for measuring outer rotation curves and low-surface-brightness structures associated with accretion and tidal interactions. Extending resolved \cii\ observations from the now-routine $z\gtrsim4$ regime to $z\sim1$ would place kinematic measurements on a consistent footing across $z=1-6$ and 
    trace low-mass galaxies out to comparable physical radii across time. This motivates substantially improved sensitivity in Bands 8--10, a frequency range uniquely accessible from ALMA's site. At $z=1-3$, low-$J$ CO remains the primary tracer of the molecular-gas surface density and kinematics, but it becomes increasingly confined to dense central regions in lower masses and higher redshift galaxies due to reduced dust shielding and metallicity, the rising CMB temperature, and increases in excitation with density/temperature \citep{daCunha_2013,glover_2016, olsen_2016,bisbas_2025, anirudh_2025}. \ci\ and \oi\ therefore become increasingly important tracers of gas surface density and flows \citep{glover_2016,bisbas_2025,parente_2026}. Joint CO/\oi, \ci, and \cii\ observations are therefore essential to connect the dynamics of dense molecular gas to the more extended cold-gas reservoir. 
    Through its tenfold sensitivity increase and full simultaneous bandwidth, ALMA2040 would extend these joint measurements down to $M_\star>10^9$\,\msun\ and SFR$\sim10$ \myr\ for unlensed galaxies ($M_\star>10^8$\,\msun\ and SFRs of a few \myr\ for lensed galaxies), enabling routine dynamical studies of Milky Way progenitors and below-main-sequence galaxies.

    \subsection{Why a tenfold sensitivity increase?}    

    \begin{description}[leftmargin=0pt,labelsep=0.5em]
    
        \item[Recovering kinematic profiles:] Cases 1--3 require $V_{\rm rot}(R)$ and $\sigma(R)$ to be well resolved. For a baryon-dominated exponential disk, the circular velocity peaks about twice the scale length, $R\approx2R_d$, so the rotation curve must be traced to $R\gtrsim4R_d$ to discern the DM contribution. Constraining the shape of the rotation curve therefore requires at least $\sim4$ independent radial measurements to $\sim4R_d \approx 2.4R_e$. At $z\sim4$, a $\mathrm{M_\star}\sim10^{9.5}$\,\msun\ main-sequence galaxy is expected to be $10\times$ fainter in \cii\ \citep{schaerer_2020} than the massive ($M_\star\sim10^{10.5}$\,\msun) galaxies currently resolved with ALMA (Fig.~\ref{fig:parameter_space}), but also $3-4\times$ smaller \citep{vdW_2014,Shibuya_2015}, yielding a $\sim3\times$ lower line brightness per resolution element. Combined with the finer spectral resolution ($\gtrsim2$\,\kms) needed to recover the $\sigma$ profiles at lower SFR, this implies a total sensitivity requirement of $\gtrsim9\times$ that of ALMA, naturally motivating the sensitivity improvement targeted by ALMA2040. A second requirement is detecting emission at the large radii where DM dominates the enclosed mass. Stacked \cii\ observations of $z\gtrsim4$ galaxies reveal low-surface-brightness emission extending well beyond the radii reached by current observations \citep[$\gtrsim10$\,kpc:][]{fujimoto_2019,novak_2020,ginolfi_2020a}, demonstrating that existing studies are limited by surface-brightness sensitivity at precisely the levels needed to probe the DM-dominated regime. The tenfold increase in line sensitivity provided by ALMA2040 would enable routine detection of \cii\ at the large radii where dark matter is expected to dominate the enclosed mass budget. 

        \item[Recovering gas mass distributions:] All science cases also require resolved measurements of the gas mass distribution, $\Sigma_{\rm gas}(R)$. For a representative $M_\star\sim10^{10}$\,\msun\ galaxy at $z\sim2$, tracing the gas distribution to $2.4 R_e$ requires sensitivity to radii of $\sim5$ kpc. As a benchmark, recent observations of \ci\ in the $z=2.4$ starburst ``W'' traced emission to similar radii at $0.12''$ resolution in 3.6\,h \citep{kaasinen_2026}. Adopting typical gas fractions and \ci-to-H$_2$ conversion factors \citep{tacconi_2020,heintz_2020,boogaard_2020} a typical $M_\star\sim10^{10}$\,\msun\ galaxy at $z\sim2$ is $\sim30\times$ fainter than W. The lower required sensitivity is partially offset by the substantially narrower line width expected for the lower mass system ($\sim150-200$\,\kms\ vs the $\sim900$\,\kms of W). Thus, resolved \ci\ observations would require $\sim700$\,h with ALMA. The tenfold line sensitivity improvement for ALMA2040 would reduce this to $\sim7$\,h, enabling resolved measurements of $\Sigma_{\rm gas}(R)$ in statistically significant samples of representative $z\sim2$ disk galaxies rather than being restricted to the most extreme gas-rich starbursts. 

        \item[Resolving interactions:] Fig.~\ref{fig:merger-vs-disk} shows how ALMA2040 will resolve a challenging scenario of a typical merger of low-mass galaxies at $z\sim6$. In this example, the tidal bridges account for 50\% of the total \cii\ emission. A 10 hour observation with current ALMA fails to detect either the faint companion or the diffuse tidal structures, whereas the same integration with ALMA2040 clearly recovers both at high significance. Moreover, ALMA2040 resolves the rotation curve of the primary galaxy (G1) well beyond the stellar disk, where constraints on the DM mass distribution are strongest. The velocity field of G1 further reveals kinematic distortions induced by tidal torques, producing a twisted velocity gradient and clear departures from regular disk rotation. A 10 hour observation therefore constrains the role of interactions in disk assembly (Case 1), recovers the rotation curve needed to model the dark matter distribution (Case 2), and directly probes the impact of gravitational torques on angular-momentum transport (Case 3). Thus, ALMA2040 can routinely deliver resolved measurements of the gas distribution, rotation curve, and velocity field of low-mass interacting/merging galaxies. 
    \end{description}

\section{Synergies with future facilities}
    \label{sec:synergies}

    \textbf{30--40m class optical/NIR facilities:} Next-generation optical telescopes, including the ELT and GMT, will provide comparable-resolution observations of the stars and ionised gas in the same $z>1$ galaxies whose cold ISM will be resolved by ALMA2040. At $z=1-3$, this synergy will -- for the first time -- enable routine comparisons of resolved cold-gas and stellar kinematics within the same galaxies, revealing how disk galaxies acquire and redistribute angular momentum through differences in the rotation, dispersion, and alignment of their gaseous and stellar components \citep{martinsson_2013,el_badry_2018,ristea_2022}. Moreover, combining resolved stellar mass distributions from ELT and GMT imaging, with ALMA2040's cold-gas kinematics and surface densities will reduce the disk-halo degeneracy while directly connecting stellar bars, spiral arms, and tidal features to the cold-gas flows that redistribute mass and angular momentum through galaxy disks. At $z>3$, obtaining resolved stellar kinematics and rest-NIR observations of low-mass galaxies will be challenging even for 30-40 m class telescopes, leaving ALMA2040 as the primary means of probing their dynamical state.
    
    \textbf{The ngVLA:} ALMA2040 and the ngVLA will provide powerful, complementary views of the cold gas reservoirs in $z>1$ galaxies (Fig.~\ref{fig:alma2040_v_ngvla}). ALMA2040 will deliver high-resolution, simultaneous observations of \cii, \ci\ and the mid- and high-$J$ CO transitions that trace excited molecular gas. It will also provide continuous coverage of both \ci\ transitions across $z=1-6$ and uniquely enable resolved studies of \cii\ and \oi. Complementary to this, the ngVLA will uniquely resolve low-$J$ CO transitions at $z>3$, providing a view of the cold, diffuse molecular gas. The combination of \cii, high-$J$ CO, and low-$J$ CO measurements will connect gas excitation, dynamics, and feedback within a single framework. 
    By complementing the ngVLA's view of cold, diffuse gas, ALMA2040 will complete the connection between diffuse gas reservoirs, cold-gas dynamics, and star formation across cosmic time.

\section{Summary}

     The next decades will transform our view of galaxy assembly across multiple phases of the baryonic cycle. Yet without a major advance in sub-/millimetre capabilities, the cold gas that fuels disk assembly and carries most of a galaxy's mass and angular momentum will remain largely unresolved. ALMA2040 is uniquely positioned to fill this gap. By extending resolved studies from rare starbursts and massive galaxies to representative galaxy populations, ALMA2040 will reveal how galaxies acquired their gas and angular momentum and when dynamically cold disks emerged. By resolving cold gas in Milky Way progenitors, ALMA2040 will provide the gas-phase counterpart to Galactic archaeology studies of the Milky Way, revealing the reservoirs and dynamics from which its stellar populations formed. By recovering cold-gas kinematics to large radii across a wide range of galaxy masses and cosmic epochs, ALMA2040 will probe the densities and profiles of DM halos during the peak epoch of galaxy assembly, providing a new test of DM models and the interplay between baryonic feedback and halo evolution. By resolving gas flows, tidal structures, and interaction-driven perturbations, it will reveal how mass and angular momentum are redistributed between and within galaxies. Together, these capabilities will provide the first statistical view of how the progenitors of present-day spiral galaxies assembled.

\section*{Acknowledgements}
This work was developed as part of the broader ALMA2040 community effort. We acknowledge the Lorentz Center and NOVA for supporting the community workshops and coordination activities that helped initiate and shape the ALMA2040 White Paper process. We also thank the two referees whose feedback substantially improved the quality of this paper. \\

\bibliographystyle{apj_mod}
{\bibliography{bibliography}}

\newpage

{
\footnotesize{
\noindent %
$^6$ Leiden Observatory, Leiden University, PO Box 9513, NL-2300 RA Leiden, The Netherlands \\ 
$^7$ Observatorio Astron\'{o}mico Nacional (IGN), C Alfonso XII 3, E-28014 Madrid, Spain, \\ 
$^8$ European Southern Observatory (ESO), Karl-Schwarzschild Str. 2, 85748 Garching, Germany \\ 
$^{9}$ Scuola Internazionale Superiore Studi Avanzati (SISSA), Physics Area, Via Bonomea 265, 34136 Trieste, Italy, \\ 
$^{10}$ INAF, Osservatorio Astronomico di Trieste, via Tiepolo 11, I-34131, Trieste, Italy \\ 
$^{11}$ Department of Physics and Astronomy, Chalmers University of Technology, SE-412 96 Gothenburg, Sweden \\ 
$^{12}$ Kavli Institute for Cosmology, University of Cambridge, Madingley Road, Cambridge CB30HA, UK \\ 
$^{13}$ Department of Physics and Astronomy and George P. and Cynthia Woods Mitchell Institute for Fundamental Physics and Astronomy, Texas A\&M University, 4242 TAMU, College Station, TX 77843-4242, USA \\ 
$^{14}$ Università di Pisa, Dipartimento di Fisica ``Enrico Fermi'', Largo Bruno Pontecorvo 3, Pisa I-56127, Italy \\ 
$^{15}$ SKA Observatory, Jodrell Bank, SK11 9FT, UK \\ 
$^{16}$ Instituto de F\'{i}sica y Astronom\'{i}a, Universidad de Valpara\'{i}so, Avda. Gran Bretaña 1111, Valparaíso, Chile \\ 
$^{17}$ Millennium Nucleus for Galaxies (MINGAL), Av. Ejercito 441, Santiago 8370191, Chile \\ 
$^{18}$ Institut d’Astrophysique de Paris, Sorbonne Universit\'{e}, CNRS, UMR 7095, 98 bis bd Arago, 75014 Paris, France \\ 
$^{19}$ INAF, Osservatorio Astrofisico di Arcetri,Largo Enrico Fermi 5, 50125, Firenze, Italy \\ 
$^{20}$ Purple Mountain Observatory, Chinese Academy of Sciences, 10 Yuanhua Road, Nanjing 210023, China \\ 
$^{21}$ State Key Laboratory of Radio Astronomy and Technology, Purple Mountain Observatory, Chinese Academy of Sciences, 10 Yuanhua Road, Nanjing 210023, China \\ 
$^{22}$ University of Ljubljana, Faculty of Mathematics and Physics, Jadranska ulica 19, SI-1000 Ljubljana, Slovenia \\ 
$^{23}$ Dipartimento di Fisica e Astronomia, Universit\'{a} di Firenze, Via G. Sansone 1, I-50019, Sesto F.no (Firenze), Italy \\ 
$^{24}$ National Radio Astronomy Observatory, 520 Edgemont Road, Charlottesville, VA 22903, USA \\ 
$^{25}$ Observatorio Astron\'{o}mico Nacional, C/ Alfonso XII 3, E-28014 Madrid, Spain \\ 
$^{26}$ Pole Scientific, University Catholic of Lyon- University of Lyon, 10 place des Archives 69288, Lyon, France \\ 
$^{27}$ David A. Dunlap Department of Astronomy and Astrophysics, University of Toronto, 50 St. George Street, Toronto, Ontario, Canada \\ 
$^{28}$ Department of Astrophysics, University of Vienna, Türkenschanzstraße 17, 1180, Vienna, Austria \\ 
$^{29}$ International Centre for Radio Astronomy Research (ICRAR), The University of Western Australia, M468, 35 Stirling Highway, Crawley,WA 6009, Australia \\ 
$^{30}$ Instituto de Estudios Astrof\'{i}sicos, Facultad de Ingenier\'{i}a y Ciencias, Universidad Diego Portales, Avenida Ej\'{e}rcito Libertador 441, Santiago, Chile. \\ 
$^{31}$ Kavli Institute for the Physics and Mathematics of the Universe (WPI), The University of Tokyo, Kashiwa, Japan \\ 
$^{32}$ Centre for Astrophysics \& Supercomputing, Swinburne University of Technology, Hawthorn, VIC 3122, Australia \\ 
$^{33}$ Institut d’Astrophysique de Paris, Sorbonne Universit\'{e}, UPMC Université Paris 6 and CNRS, UMR 7095, 98 bis boulevard Arago, F-75014 Paris, France \\ 
$^{34}$ Université de Strasbourg, CNRS, Observatoire astronomique de Strasbourg, UMR 7550, F-67000 Strasbourg, France \\ 
$^{35}$ Instituto de F\'isica Fundamental, CSIC, Calle Serrano 123, 28006 Madrid, Spain \\ 
$^{36}$ Sub-department of Astrophysics, Department of Physics, University of Oxford, Denys Wilkinson Building, Keble Road, Oxford, OX1 3RH, UK \\ 
}
}

\end{document}